\documentclass[aps,prl,twocolumn,a4paper,10pt,notitlepage,footinbib,superscriptaddress,showpacs]{revtex4-1}%
\usepackage[english]{babel}
\usepackage[T1]{fontenc}
\usepackage{endnotes}
\usepackage{amssymb,amsmath,amsfonts}
\usepackage{textcomp}
\usepackage{graphicx,color}
\usepackage[utf8x]{inputenc}
\usepackage{float}
\usepackage{placeins}
\usepackage{multirow}
\usepackage{colortbl}
\usepackage{tabulary}
\usepackage{etoolbox}

\begin{document}

\title{Rotation and propulsion in 3d active chiral droplets}

\author{Livio Nicola Carenza}
\affiliation{Dipartimento  di  Fisica,  Universit\`a  degli  Studi  di  Bari  and  INFN,  via  Amendola  173,  Bari,  I-70126,  Italy}

\author{Giuseppe Gonnella} 
\affiliation{Dipartimento  di  Fisica,  Universit\`a  degli  Studi  di  Bari  and  INFN,  via  Amendola  173,  Bari,  I-70126,  Italy}

\author{Davide Marenduzzo}
\affiliation{SUPA, School of Physics and Astronomy, University of Edinburgh, Mayfield Road, Edinburgh EH9 3JZ, United Kingdom}

\author{Giuseppe Negro}
\affiliation{Dipartimento  di  Fisica,  Universit\`a  degli  Studi  di  Bari  and  INFN,  via  Amendola  173,  Bari,  I-70126,  Italy}

\begin{abstract}
Chirality is a recurrent theme in the study of biological systems, in which active processes are driven by the internal conversion of chemical energy into work. Bacterial flagella, acto-myosin filaments and microtubule bundles are active systems which are also intrinsically chiral. 
Despite some exploratory attempt to capture the relations between chirality and motility, no intrinsically chiral system has ever been analyzed so far.
To address this gap in knowledge, here we study the effects of internal active forces and torques on a three-dimensional droplet of cholesteric liquid crystal (CLC) embedded in an isotropic liquid. We consider tangential anchoring of the liquid crystal director at the droplet surface. Contrary to what happens in nematics, where moderate extensile activity leads to droplet rotation, cholesteric active droplets exhibit a lot more complex and variegated behaviors. 
We find that extensile force dipole activity stabilises complex defect configurations whose orbiting dynamics couples to thermodynamic chirality to propel screw-like droplet motion. Instead, dipolar torque activity may either tighten or unwind the cholesteric helix, and if tuned can power rotations with an oscillatory angular velocity of zero mean. 
\end{abstract}

\maketitle







Chirality is a generic feature of most biological matter~\cite{ramaswamy2010,Zhou1265,naganathan2014}. A right-left asymmetry may arise at either the microscopic or macroscopic level, and be due to thermodynamic (passive) or non-equilibrium (active) effects. For instance, a microtubule-motor mixture breaks chiral symmetry in two ways. First, microtubules are intrinsically helical~\cite{zhang16}. Second, kinesin or dynein motors exploit ATP hydrolisis to twist their long chains and apply a nonequilibrium active torque on the fibres they walk along~\cite{Wang2012}. Similarly, bacteria such as~\emph{Escherichia coli}, but also sperm cells, are equipped with long helical flagella. Motor proteins anchored to the cellular membrane generate torques to impart rotational motion on the flagella, whose helix generates a flow in the viscous environment leading to cell propulsion~\cite{Purcell1997,Riedel2005}.

Biological fluids are \emph{active}, as they are internally driven by the constant injection of energy, which prevents them from relaxing towards any thermodynamic steady state~\cite{marc2013}. A simple and successful theory to model an active fluid is to  approximate each of its microscopic constituents (e.g., a microtubule or a single bacterium) as an entity which exerts a dipolar force on the environment~\cite{hatwalne2004}. The dipole direction introduces orientational order resembling that of liquid crystals (LC). The resulting \emph{active nematic} has been found, both experimentally and numerically, to develop unexpected and striking behaviors such as \emph{spontaneous flow}, \emph{active turbulence} and \emph{superfluidic} states~\cite{cates2008,reviewYeomans}. Much effort has been spent in the last decades to capture the essential dynamics of active fluids, ranging from multi-particle models to continuum theories~\cite{digregorio2018,bonelli2019,negro2018}.  The theory has been able to capture experimentally observed features like turbulent-like patterns in $2d$ bacterial films~\cite{wensink2012}, or spatiotemporal pattern formation and topological defect dynamics in active emulsions containing microtubule-kinesin mixtures~\cite{sanchez2012}.

Understanding the outcome of the interplay between chirality and activity is an important and timely question. In stark contrast with the case of achiral active nematics, which has commanded a lot of attention in recent years, much less is known about the dynamics of chiral active systems~\cite{tjhung2017,maitra2019,whitfield2017}. Previous work has mainly focused on cases where chirality only enters the system because of activity, in the form of a nonequilibrium torque dipole~\citep{tjhung2017,maitra2019}. Instead, we consider here a system which is {\it inherently} chiral and apolar, and so can be modelled -- in the passive phase -- as a cholesteric liquid crystal (CLC)~\cite{zumer,FrankPrice}. 
Specifically, here we study a $3d$ active CLC droplet with tangential orientation of the director at its surface.  
In this setup, an active nematic droplet can only sustain uniform rotational motion, driven by bend deformations localised around the equatorial circle of the droplet (Fig.~1). Instead, an intrinsically chiral droplet displays a much richer dynamical behaviour. First, we find that a force dipole activity enables a new motility mode, where the rotational motion of the surface defects is converted into propulsion. This mechanism requires chirality to reconfigure the pattern of surface defects. It is not possible in a nematic, 
where the symmetry in defect position prevents any translational motion. 
Second, a torque dipole activity sets up a sustained mirror rotation of two pairs of disclinations which periodically adsorb onto and depin from the droplet surface. Again, no such state can be found in an originally nematic system. We also characterise how the active flow and orientation patterns evolve as the ratio between the droplet size and pitch increases.


The ordering properties within the CLC droplet are described, in the uniaxial limit, by the nematic tensor $\mathbf{Q}= S(\mathbf{n}\mathbf{n}-\mathbf{I}/3) $, where the \emph{director} field $\mathbf{n}$ is a unit head-less vector describing the local average orientation of the components and $S$ is a scalar field expressing the degree of order. The equilibrium properties are derived by a suitable Landau-De Gennes free energy functional, ${\mathcal F}$, of which key parameters are the equilibrium cholesteric pitch $p_0$ and the ratio between the droplet diameter and the pitch, $N\equiv 2R/p_0$, which also counts the number of windings of the director within the droplet. Activity is introduced through a coarse grained description of force and torque dipoles~\cite{ghose2014}. These result in a non-equilibrium stress tensor that can be respectively expressed as $\sigma_{\alpha \beta}^{af}= -\zeta \phi Q_{\alpha \beta}$ (force dipoles) and $\sigma_{\alpha \beta}^{at}= -\bar{\zeta} \epsilon_{\alpha \beta \mu} \partial_\nu (\phi Q_{\mu \nu}) $ (torque dipoles), where $\phi$ is the concentration of active material, $\epsilon_{\alpha \beta \mu}$ the Levi-Civita tensor, while $\zeta$ and $\bar{\zeta}$ are proportional to the strength of active force and torque dipoles respectively.
Positive values of $\zeta$ correspond to extensile force dipoles, whereas if $\zeta<0$ the force dipole activity is contractile. For active torque dipoles, a positive value of $\bar{\zeta}$ corresponds to an outward pair of torques, similar to that used to open a bottle cap. This is, for instance, at the base of the motility of some flagellate bacteria, such as \emph{E. Coli}: in this case the body rotate clockwise, while the flagella rotate anti-clockwise. Conversely, ${\bar{\zeta}<0}$ correspond to an inward torque pair, similar to that used to close a bottle cap. 
Unless otherwise stated, here we restrict for concreteness to the case where $\zeta>0$ and $\bar{\zeta}<0$.



\section*{Results}

\subsection*{Nematic droplet with active force dipoles: spontaneous rotation}

\begin{figure}[h!]
\centering
\includegraphics[width=1.\columnwidth]{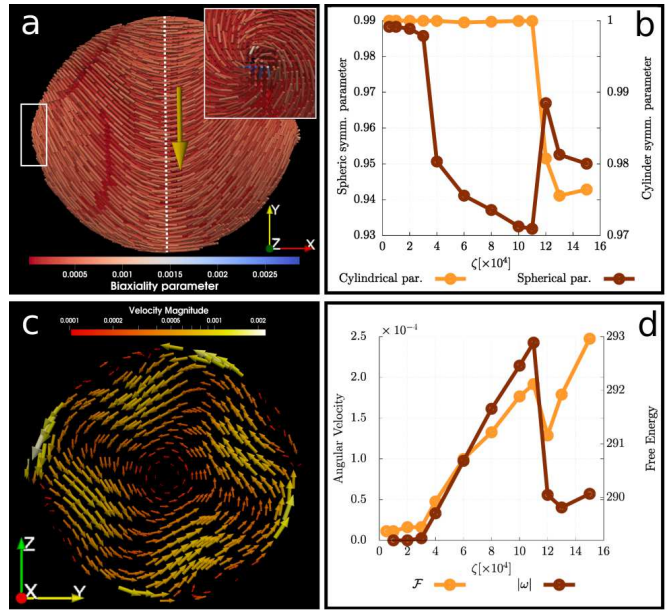}
\caption{\textit{Active nematic droplet.} 
Panel~(a) shows the configuration of the director field $\mathbf{n}$ on the droplet surface for the case in the rotational regime at $\zeta=10^{-3}$. Two stationary $+1$ boojums are formed at antipodal points in the $x$ direction. 
The inset in panel~(a) shows the director field in proximity of the boojum framed with a white box.
Bending deformations occur transversally to the long axes of the droplet and generate an active force in the direction of the yellow arrow, thus powering rotational motion in the $yz$ plane. Panel~(c) shows the velocity field on the equatorial cross section of the droplet, depicted with a dashed white line in panel~(a). The flow, induced and sustained by energy injection due to the bending deformations in the nematic pattern, exhibits four-fold symmetry.
In panel~(b) the spherical and cylindrical deformation parameters (see Material and Methods for their definitions) are used to characterize the transition from the quiescient state first to the rotational regime and then to the chaotic regime.  Analogously panel~(d) shows the angular velocity and the free energy as the activity parameter $\zeta$ is varied.
}
\label{fig:fig1}
\end{figure}

\begin{figure*}[t]
\centering
{\includegraphics[width=1.\textwidth]{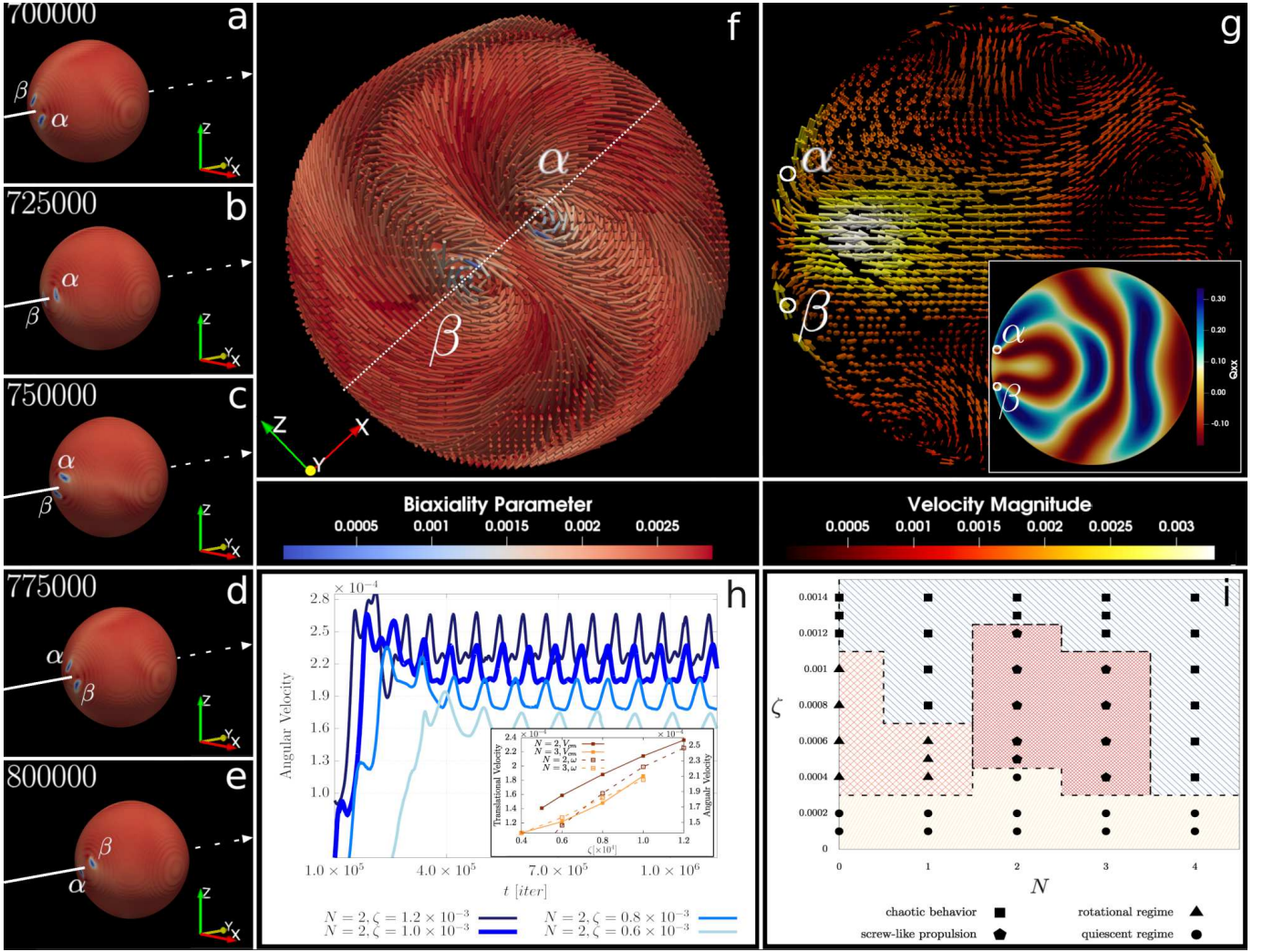}}
\caption{\textit{Screw-like propulsion in a chiral droplet with active force dipoles.} Panels (a-e) show snapshots at different times of a chiral active droplet for the case at $N=2$ and $\zeta=10^{-3}$. The contour-plot of the biaxility parameter on the droplet surface serves to identify the position of the two $+1$ defects, labelled with Greek letters $\alpha$ and $\beta$, whose configuration can be appreciated by looking at panel (f). The screw-like rotational motion generates a strong velocity field in the interior of the droplet in proximity of the two defects. The velocity field is plotted in panel (g) on a plane transversal to the plane of rotation of the two defects (dashed line in panel~(f)). The inset shows the contour plot, on the same plane, of the $Q_{xx}$ component of the Q-tensor, exhibiting an arrangement similar to the \emph{radial spherical structure}.  Panel (h) shows the time evolution of the angular velocity of the droplet for some values of $\zeta$. The inset shows the mean angular velocity and the translational velocity of the droplet as a function of $\zeta$ both for $N=2$ and $N=3$. Panel (i) summarizes the droplet behavior as a function of $\zeta$ and $N$.}
\label{fig:fig2}
\end{figure*}

\begin{figure*}[ht]
\centering
{\includegraphics[width=1.\textwidth]{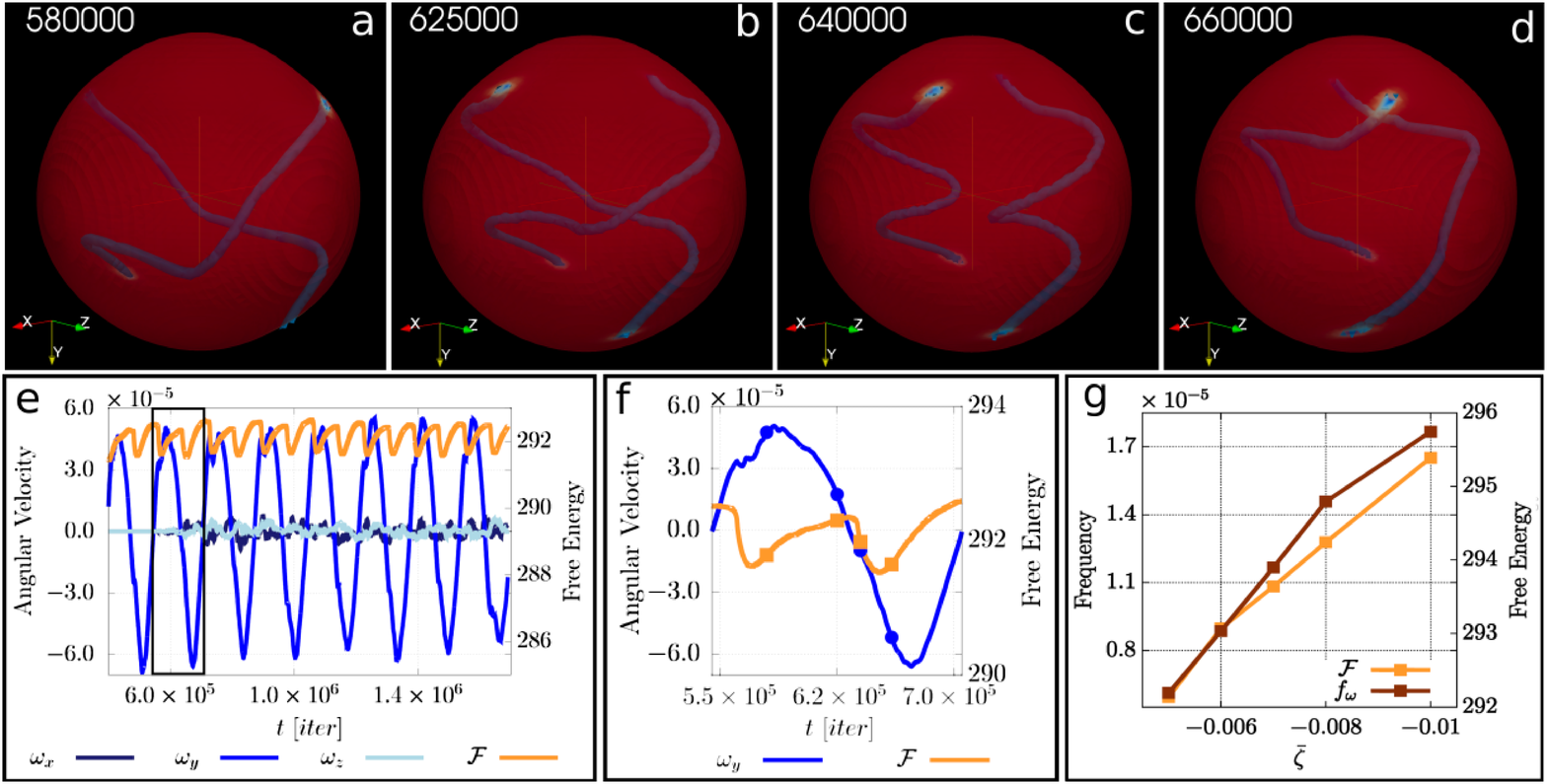}}
\caption{\textit{Disclination dance in a chiral droplet with active torque dipoles.} Panels (a-d) show snapshots of the droplet and the disclination lines for the case at $N=1$ and $\bar{\zeta}=-5\times 10^{-3}$. The four $+1/2$ defects rotate in pairs in opposite directions (top defects rotate anti-clockwise, while bottom defects rotate oppositely). As the defects rotate the two disclination lines first create a link (b), then they recombine (c) and finally relax into a configuration close to the initial one (a) but rotated. The angular velocity, null on average, oscillates from positive to negative values as shown in panel (e). Here the time evolution of the free energy shows that $\mathcal{F}$ oscillates with a double frequency. Inset shows the behavior of $\omega_y$ and $\mathcal{F}$ in the region framed with the black box. Marked dots here denote  the points corresponding to the snapshots.}
\label{fig3}
\end{figure*}

We start by describing the dynamics of a nematic droplet (infinite pitch, or $N=0$), with active force dipoles -- this is a useful limit to which the chiral results will be compared. In this system, the main control parameter is the dimensionless activity $\theta=\zeta R^2/K$, which measures the relative contribution of activity and elasticity (see Materials and Methods, where we also provide a possible mapping between simulation and physical units). The active nematic case has been previously studied in the literature~\cite{activenematicdroplet}, although never for tangential anchoring in $3d$.


As the dimensionless activity $\theta$ is increased, we observe three possible regimes. For low values of $\theta$  the droplet is static, and the director field attains a boojum-like pattern, with two antipodal surface defects of topological charge $+1$. This is one of the patterns which can be found in a passive nematic droplet with tangential anchoring, and satisfies the hairy-ball theorem~\cite{Poincaretheorem} which states that the sum of the topological charges of a vector field tangential to a closed surface is equal to its Euler characteristic (which is $+2$ for a sphere). As activity is increased, this \emph{quiescent phase} gives way to another regime, where the droplet spontaneously rotates in steady state (Fig.~\ref{fig:fig1}, "SI Appendix, Movie~1"). The quiescent droplet has spherical symmetry, whereas it deforms in the rotating phase, attaining the shape of a prolate ellipsoid of revolution (the asphericity parameter though shows only up to $\sim 10\%$ variation from a sphere, Fig.~\ref{fig:fig1}a,b). The director field on the droplet surface exhibits bending deformations, typical of extensile suspensions~\cite{ramaswamy2010,marc2013}, that are strongest at the equator. They act as a momentum source (see yellow arrow in Fig.~\ref{fig:fig1}a), hence power stable rotational motion.
The
flow has the pattern of a single vortex inside the droplet, which is stronger close to the surface. The rotational flow exhibits four-fold symmetry in the equatorial plane, and is compensated by a counteracting velocity field outside the droplet -- ensuring overall angular momentum conservation. The rotational velocity, $\omega$, is linearly proportional to the activity $\zeta$. This scaling can be rationalised by dimensional analysis, or by equating the torque per unit volume introduced by activity, which should scale as $\zeta$, to the one which is dissipated, given by $\gamma_1 \omega$, where $\gamma_1$ is the rotational viscosity. 

For still larger $\theta$, the droplet rotates and moves in a chaotic manner. This regime is the droplet analogue of what is known as \emph{active turbulence}~\cite{activeturbulence,copar2019}-- the chaotic dynamics observed in an active nematic fluid. In the chaotic regime, motion is random and the cylindrical symmetry of the droplet shape is lost (Fig.~\ref{fig:fig1}b). Defect dynamics on the surface is also erratic, and we observe the nucleation of additional defects, not present in the quiescent or rotating regimes. These defects are topologically the ends of disclination lines which often depin from the surface and pierce the interior of the droplet ("SI Appendix, Movie~2"). The onset of the chaotic regime is due to the fact that the energy coming from activity can no longer be dissipated by a regular rotation, but is used up to generate additional defects on the surface. As the chaotic regime sets in -- characterized by the loss of cylindrical symmetry -- concurrently the free energy of the system decreases (Figs.~\ref{fig:fig1}d,f), signalling that the shape change is thermodynamically favoured. The subsequent increase in ${\mathcal F}$ at larger $\theta$ is due to defect nucleation. 






\subsection*{Cholesteric droplet with active force dipoles: screwlike propulsion}

We now consider the case of a cholesteric droplet, still with active force dipoles only. We consider the case where the twist in equilibrium is right-handed (the left-handed case is mirror symmetric). The two key control parameters are now $\theta$ and $N$. For a fixed value of $N$, increasing $\zeta$ again leads to three possible regimes, as in the nematic limit. For sufficiently large cholesteric power (e.g., $N=2$, Fig.~\ref{fig:fig2}), the first active regime encountered is, however, fundamentally different from the rotating phase of active nematics. Now the surface defect pattern is a pair of nearby $+1$ defects, reminiscent of a \emph{Frank-Price structure} 
which is seen in passive cholesterics, but only with much larger $N$ ($N\ge 5$~\cite{FrankPrice}). The  configuration of director field which we observe is known as \emph{radial spherical structure}~\cite{zumer,depablo}, with some additional distortions in the cholesteric layers due to activity (as suggested by the inset in Fig.~\ref{fig:fig2}g that gives an insight into the cholesteric arrangement in the interior of the droplet). There is a suggestive analogy between this structure and a magnetic monopole -- representing the radial orientation of the helical structure at the droplet centre -- with its attached \emph{Dirac string}~\cite{Lavrentovich,depablo}, joining the centre of the droplet with the defect pair.
In our simulations the latter represents the  region of maximal layer distortion and energy injection, as suggested by the intensity of the velocity field, plotted in Fig.~\ref{fig:fig2}g.

The two surface defects rotate around each other (Fig.~2a-e): as they do so, 
the droplet undergoes a global rotation with oscillating angular velocity (Fig.~\ref{fig:fig2}h and "SI Appendix, Movie 3"). Remarkably, this time the rotation is accompanied by a translation along the direction of the rotation axis -- thereby resulting in a 
screwlike motion, with the axis of the rototranslation parallel to the \emph{Dirac string}.
This motility mode is compatible with the chiral symmetry of the system, which introduces a generic non-zero coupling between rotations and translations. 
The strong deformations induced by the two nearby rotating $+1$ surface defects are responsible for the intense flow that develops inside the droplet and is maximum at the rear (Fig.~\ref{fig:fig2}g). This active flow is the one powering propulsion. The symmetry of the flow corresponds to that of a macroscopic pusher.
Mechanistically, therefore, activity is required to power droplet rotation, and chirality is needed to couple rotation to motion. As the motion is screwlike, the linear and the angular velocity are proportional to each other -- a similar argument to that used for active nematics also shows that they should both scale approximately linearly with $\theta$, and we found this to hold for our simulations (Fig.~\ref{fig:fig2}h, inset). 

A phase diagram in a portion of the $(N,\zeta)$ plane is shown in Fig.~\ref{fig:fig2}i.
The results -- which are independent of the (random) initial conditions for the parameter range explored here-- show that for small activity the droplet sets into a quiescent regime indipendently of the cholesteric power. This state is characterized by weak bending deformation of the LC network on the droplet surface, which are not enough to power any self-sustained motion. As activity is increased different behaviors arise: for weak cholesteric power ($N \leqslant 1$) stationary rotational motion sets up, while screwlike propulsion needs the defects to relocate to one hemisphere creating a dipolar pattern. This is found to be only possible for a limited range of $\zeta$ and only for $N=2,3$. Indeed, at higher cholesteric power ($N \geqslant 4$), the droplet sets into the chaotic phase even at intermediate activity, a regime characterized by defect nucleation and disordered droplet motility that can be found at any $N$ for sufficiently large values of $\zeta$.

\subsection*{Cholesteric droplets with active torque dipoles: rotation and disclination dance}

We next consider the case of a cholesteric droplet with active torque dipoles. These are able to introduce a nonequilibrium twist in a nematic droplet~\cite{elsen}, whose handedness may reinforce or oppose the handedness of the thermodynamic twist, which is determined by $q_0$. The strength of the nonequilibrium twist can be measured by the dimensionless number $\bar{\theta}=|\bar{\zeta}|R/K$, whilst that of the equilibrium one can be assessed by $N$.

We find that the most interesting dynamics, in the case of a right-handed equilibrium twist ($q_0>0$), occurs for $\bar{\zeta}<0$ (inward torque dipole, leading to a conflict between the nonequilibrium and equilibrium twist). In this situation, for $N=1$, we find that the droplet is pierced by two disclination lines which end in $+1/2$ surface defects at $\bar{\zeta}=-5 \times 10^{-5}$. The droplet regularly alternates opposite sense rotations, along $\pm \hat{y}$, which are tightly regulated by the disclination dynamics (Fig.~\ref{fig3} and "SI Appendix, Movie 4"). 
The helical axis is here approximately parallel to $\hat{z}$, with the director almost parallel to $\hat{x}$ in the centre of the droplet.
At the beginning of the rotation cycle shown in Fig.~\ref{fig3}, the disclinations wind once around each other in a right-handed fashion. Equivalently, if we were to orient both the disclinations along the positive $\hat{y}$ axis, we can associate the single crossing visible in the projection of Fig.~\ref{fig3}a with a positive writhe~\cite{bates2005dna} (as the top disclination can be superimposed on the bottom one via an anticlockwise rotation).  As the system evolves, due to the internal torque dipoles, the pair of surface defects in the top hemisphere rotates counterclockwise, while that in the bottom hemisphere rotates clockwise (Fig.~\ref{fig3}b). This motion increases the winding of the disclinations, until they rewire to form two separate right-handed helices  (Fig.~\ref{fig3}c -- if we were to extend the two disclinations along $\hat{z}$, they would be unlinked). The regular switches in the sense of droplet rotation beat the time of the  disclination dance visualised in Fig.~\ref{fig3}a-d. Rotation inversion occurs just at the time when the defect rewiring happens, as a result of the top/bottom asymmetry in the disclination configuration. Becase the surface defects are the regions of strongest elastic deformation, their location determines the vortical flow pattern. Upon rewiring, the relative contributions of the defects in the top and bottom hemispheres change, and this dictates the change in sense of rotation. We find that the evolution of the angular velocity mirrors that of the overall free energy of the system, with a small time delay: we argue that this is because the stress stored in the elastic deformations plays a large role in powering the motion. 
The frequency of the free-energy oscillation is twice that of the angular velocity $f_w$ (see panels~(e-f)), a behavior in line with the fact that configurations in panel~(a) and~(d) are specular with respect to the rotation plane and energetically equivalent.

Unlike in the active nematic case, where rotation is powered by force dipoles, here the dynamics is driven by torque dipoles. The different physics leaves a signature in the scaling of the (maximal) angular velocity, which now can be estimated as $|\bar{\zeta}|/R$ from dimensional analysis. We confirmed this behavior by simulating cholesteric droplets of different radius, ranging from $R=18$ to $32$, keeping fixed the  pitch of the cholesteric helix $p_0=64$.

The properties of a cholesteric droplet fueled by torque dipoles is highly sensitive to both the active doping and the twisting number $N$. Indeed, the dynamics described so far at $N=1$ is stable only for a limited range of activity ($ 5 \times 10^{-3} \leqslant |\bar{\zeta}| \leqslant 12 \times 10^{-3}$).
Small values of $|\bar{\zeta}|$ ($<5 \times 10^{-3}$), are not enough to excite the splitting of the two boojums and generate instead bending deformations of the LC pattern at $N=1$, similar to those shown in Fig.~\ref{fig:fig1}a. In this case the droplet sets into a stationary rotational motion characterized by small angular velocity ($|\omega|\sim \mathcal{O}(10^{-6})$)~("SI Appendix, Fig.~1"). 
If activity exceeds a critical threshold, $|\bar{\zeta}| > 12 \times 10^{-3}$, nucleation of further defects on the droplet surface leads to droplet deformation with consequent chaotic dynamics~("SI Appendix, Fig.~S1").
The competition between active and equilibrium chirality has important effects when $N$ is changed. Indeed, a further key dimensionless number to determine the behaviour of a cholesteric droplet with active torque dipoles is $\bar{\zeta}/(q_0 K)$, or equivalently, the ratio between the pitch and the ``active torque length'' $K/\bar{\zeta}$. The latter can be thought of as the nonequilibrium pitch, or the modulation in twist due to the action of the active flow. We would then expect that for larger $q_0$ (i.e., larger $N$ at fixed $R$), a rotating regime as in Fig.~\ref{fig3} can be obtained by increasing $\bar{\zeta}$.
Our simulations confirm that the range of stability of stationary rotation widens as $N$ is increased, while the set up of the mirror rotation regime moves towards more intense $|\bar{\zeta}|$. Nevertheless, if $N \geqslant 4 $, the droplet directly moves from the rotational to the chaotic regime, analogously to what happens in a cholesteric droplet fueled by force dipoles only.

It is notable that the disclination dance which we observe at intermediate $|\bar{\zeta}|$ is also reminiscent of that seen experimentally in active nematic shells~\cite{Keber1135,zhang16,activenematicdroplet} made up of microtubule-molecular motor mixtures. Despite the different geometry, our results suggest that active torque dipoles can provide an alternative mechanism powering rotation.






\FloatBarrier

\section*{Discussion and Conclusions}

In this work, we have analysed the hydrodynamic instabilities of an active chiral nematic droplet with tangential anchoring. We introduced activity as either a collection of dipolar forces or torques. Our simulations show that the interplay between activity and thermodynamic chirality in a $3d$ fluid droplet leads to a strikingly rich phenomenology. This includes screwlike droplet motion -- for dipolar active forces -- and global rotation with periodic sense inversion -- for dipolar active torques. 

Screw-like motion arises due to the coupling between thermodynamic chirality and a rotational flow, which is powered by extensile dipolar force activity due to spontaneous bend deformation at the droplet surface and its interior, compatible with tangential anchoring. This motility mode is therefore a rototranslation that is similar to that performed by a helical propeller. For a fluid with active torque dipoles, instead, global rotations with intermittent sense arise when the active torque favours a different twist with respect to that introduced by the thermodynamic chirality. Here the rotation is coupled to the rotation of helical disclination lines which pierce the droplet interior. 
For both screw-like and intermittent rotation, the surface defects arising due to tangential anchoring play a fundamental role. For the former phenomenon, defect rotation -- induced by activity -- is converted into translatory motion due to the underlying chirality. For the latter, disclination rewiring determine the change in rotation sense. The mechanisms underlying 
motility regimes are therefore defect-dependent, and qualitatively different from those analysed in~\cite{elsen}, which occur in defect-free droplets, but are associated with large deformation of the droplet shape.


It is of interest to think of the generic models developed in this work with respect to the dynamics of self-motile and rotating living active gels, which are found in bacterial and eukaryotic cells. In both cases, the cytoplasm includes chiral cytoskeletal filaments, composed of either MreB or actin, which are dynamical helical fibre and which, in the absence of any activity, would self-assemble into cholesteric phases. Additionally, molecular motors walking on such helical fibres will create active forces and torque dipoles, either in the bulk or in a cortex close to the surface. In some cases, such as that of {\it Spiroplasma} bacteria, or of single-cell parasites~\cite{spiroplasma,trypanosome,toxoplasma}, screw-like motility is observed. This has often been associated to the twisting or rotation of cytoskeletal filaments, which generates translatory motion, so that the underlying mechanism is that of a helical propeller, as in our droplets in Fig.~\ref{fig:fig2}. Our results, together with those previously reported in~\cite{elsen}, show that there are multiple motility modes which chiral active microorganism may employ, and it would be of interest to look more closely for analogues of these in nature.

Our novel \emph{active cholesteric droplets} may also be realised in practice by self-assembling active liquid crystalline droplets synthetically (a possible mapping between numerical and physical units has been provided in the
\emph{Material and Methods} section, based on similar (nematic) experimental systems~\cite{activenematicdroplet,sanchez2012}).
This could be done, for instance, by using active nematics with a chiral dopant~\cite{sanchez2012}. Although these systems usually form shells on the interface of an oil-water emulsion, we expect that a sufficiently thick shell would behave in a qualitatively similar way to our droplets. In these active liquid crystal shells, anchoring of the director field (the microtubule orientation in~\cite{sanchez2012}) is tangential as in our droplets, so defect topology should play a key role for both systems. Another potential candidate system is a cholesteric DNA or chromatin globule interacting with molecular motors~\cite{activechromatin,activechromatin1}.

\section{Material and Methods}
\label{sec:mat&met}

We considered an incompressible fluid with mass density $\rho$ and divergence-free velocity $\mathbf{v}(\mathbf{r},t)$.
A scalar, globally conserved concentration field $\phi(\mathbf{r},t)$ and the $\mathbf{Q}$-tensor respectively account for the concentration of active material and its orientational order.
The equilibrium properties are described by means of the Landau-De Gennes free energy functional:
\begin{multline}
\mathcal{F}\left[\phi,Q_{\alpha \beta}\right] = \int dV \ \left[ \dfrac{a}{4} \phi^2 (\phi-\phi_0)^2 + \dfrac{k_\phi}{2} (\nabla \phi)^2
 \right. \\ \left.
+ A_0 \left[ \dfrac{1}{2} \left(1 - \dfrac{\chi(\phi)}{3} \right)\mathbf{Q}^2 -  \dfrac{\chi(\phi)}{3} \mathbf{Q}^3 +  \dfrac{\chi(\phi)}{4} \mathbf{Q}^4 \right] 
 \right. \\ \left.
+ \dfrac{K_Q}{2} \left[ (\nabla \cdot \mathbf{Q})^2 + (\nabla \times \mathbf{Q} + 2 q_0 \mathbf{Q})^2 \right]
+W (\nabla \phi) \cdot \mathbf{Q} \cdot (\nabla \phi) 
 \right],
 \label{eqn:freeE}
\end{multline}
where the constants $a$,$k_\phi$ define the surface tension and the interface width among the two phases, whose minima are found in $0$ and $\phi_0$. The liquid crystal phase is confined in those regions where $\chi(\phi)=\chi_0 + \chi_s \phi > 2.7$, with $\chi_0= 10 \chi_s = 2.5$. 
The gradient terms in $K_Q$ account for the energy cost of elastic deformations in the one-constant approximation, while $|q_0|=2\pi/p_0$, where $p_0$ is the pitch of the cholesteric helix ($q_0>0$ for right-handed chirality).
Tangential anchoring is obtained for $W<0$.
The dynamical equations governing the evolution of the system are: (i) a convection-diffusion equation for $\phi$
\begin{equation}
\partial_t \phi + \nabla \cdot (\phi \mathbf{v}) = \nabla \cdot \left( M \nabla \dfrac{\delta \mathcal{F}}{\delta \phi} \right),
\label{eqn:conv-diff}
\end{equation}
where $M$ is the mobility parameter;
(ii) the Beris-Edwards equation for the $\mathbf{Q}$-tensor:
\begin{equation}
(\partial_t + \mathbf{v}\cdot \nabla) \mathbf{Q} - \mathbf{S}(\nabla \mathbf{v},\mathbf{Q}) = \Gamma \mathbf{H},
\label{eqn:BerisEdwards}
\end{equation}
where $\mathbf{S}(\mathbf{W},\mathbf{Q})$ is the strain-rotational derivative (see "SI Appendix"), $\Gamma$ is the rotational viscosity and 
$ \mathbf{H} = - \frac{\delta \mathcal{F}}{\delta \mathbf{Q}} + \frac{\mathbf{I}}{3} Tr \left( \frac{\delta \mathcal{F}}{\delta \mathbf{Q}} \right)$ is the molecular field;
(iii) and the Navier-Stokes equation:
\begin{equation}
(\partial_t + \mathbf{v} \cdot \nabla ) \mathbf{v} = \nabla \cdot \left[ \mathbf{\sigma}^{pass} + \mathbf{\sigma}^{act} \right].
\label{eqn:Navier_Stokes}
\end{equation}
We split the stress tensor contribution into a passive and an active term. The first one accounts for the dissipative/reactive contributions (see "SI Appendix"), while 
the active stress tensor is given by:
\begin{equation}
\sigma_{\alpha \beta}^{act} = -\zeta \phi Q_{\alpha \beta} - \bar{\zeta} \epsilon_{\alpha \mu \nu} \partial_\mu (\phi Q_{\nu \beta}).
\end{equation}

The dynamical equations have been integrated by means of a  hybrid lattice Boltzmann method starting from random initial configurations of the confined LC.
The spherical and cylindrical deformation parameters have been computed as $d_{min}/d_{max}$ and $d_{min}/d_{med}$ respectively, with $d_{max} \geqslant d_{med} \geqslant d_{min}$ the time-averaged eigenvalues of the positive-definite Poinsot matrix associated to the droplet.
More details on the methods to compute the angular velocity and the degree of biaxiality of the LC are discussed in "SI Appendix".

Mapping simulation units to physical ones is non-trivial as we do not yet accurately know hydrodynamic parameters for most active matter systems. Nevertheless, by following previous studies~\cite{bonelli2019,tjhung2012}, an approximate mapping -- for an active gel made up of cytoskeletal filaments and molecular motors -- can be obtained by equating one length, time and force simulation units to respectively $L=1\mu m$, $\tau=10ms$ and $F=1000nN$. With this choice, $\zeta=10^{-4}$ corresponds to an active stress of $1$ $kPa$, whilst the typical active lengthscale $l_a\equiv{\sqrt{K/\zeta}}$ and timescale $\tau_a\sim \frac{\eta}{\zeta}$ are (for our choice of $K$ and $\eta$) $\sim 10$ $\mu m$ and $\sim 1 s$ respectively. More details, and a Table to convert between simulations and physical units, are given in the "SI Appendix".

\if{
\begin{table}[ht]
\centering
{\begin{tabular}{@{}|l|l|l|@{}}
\hline
\multicolumn{1}{|c|}{\cellcolor[HTML]{656565}{\color[HTML]{FFFFFF}Model parameters }}   &
\multicolumn{1}{|c|}{\cellcolor[HTML]{656565}{\color[HTML]{FFFFFF}Simulation units}}   &
\multicolumn{1}{|c|}{\cellcolor[HTML]{656565}{\color[HTML]{FFFFFF}Physical units}}   \\\hline
Shear viscosity, $\eta$           & $5/3$  & $1.67 \ \text{KPas}$       \\
Elastic constant, $K_Q$           & $0.01$  & $50 \ \text{nN}$      \\
Diffusion constant, $D=Ma$        & $0.007$  & $0.06\ \text{$\mu$} \text{m}^2 \text{s}^{-1}$     \\
Activity, $\zeta$    & $0-0.01$  & $(0-100) \ \text{KPa}$       \\ 
\hline
\end{tabular}\caption{Mapping between simulations units and physical units.}\label{tabel_units}}
\end{table}
}\fi 


\paragraph{Acknowledgement} Simulations have been performed at Bari ReCaS e-Infrastructure funded by MIUR through the program PON Research and Competitiveness 2007-2013 Call 254 Action I and at ARCHER UK National Supercomputing Service (http://www.archer.ac.uk) through the program HPC-Europa3. We thank E. Orlandini and A. Tiribocchi for very useful discussions.

\bibliographystyle{unsrt}
\bibliography{refs}

\newpage

\section{SI Appendix}
\section*{Active torque dipoles}
We discuss here additional results concerning the properties of a cholesteric droplet fueled by torque dipoles.  

\begin{figure*}[hb]
\centering
{\includegraphics[width=1.\textwidth]{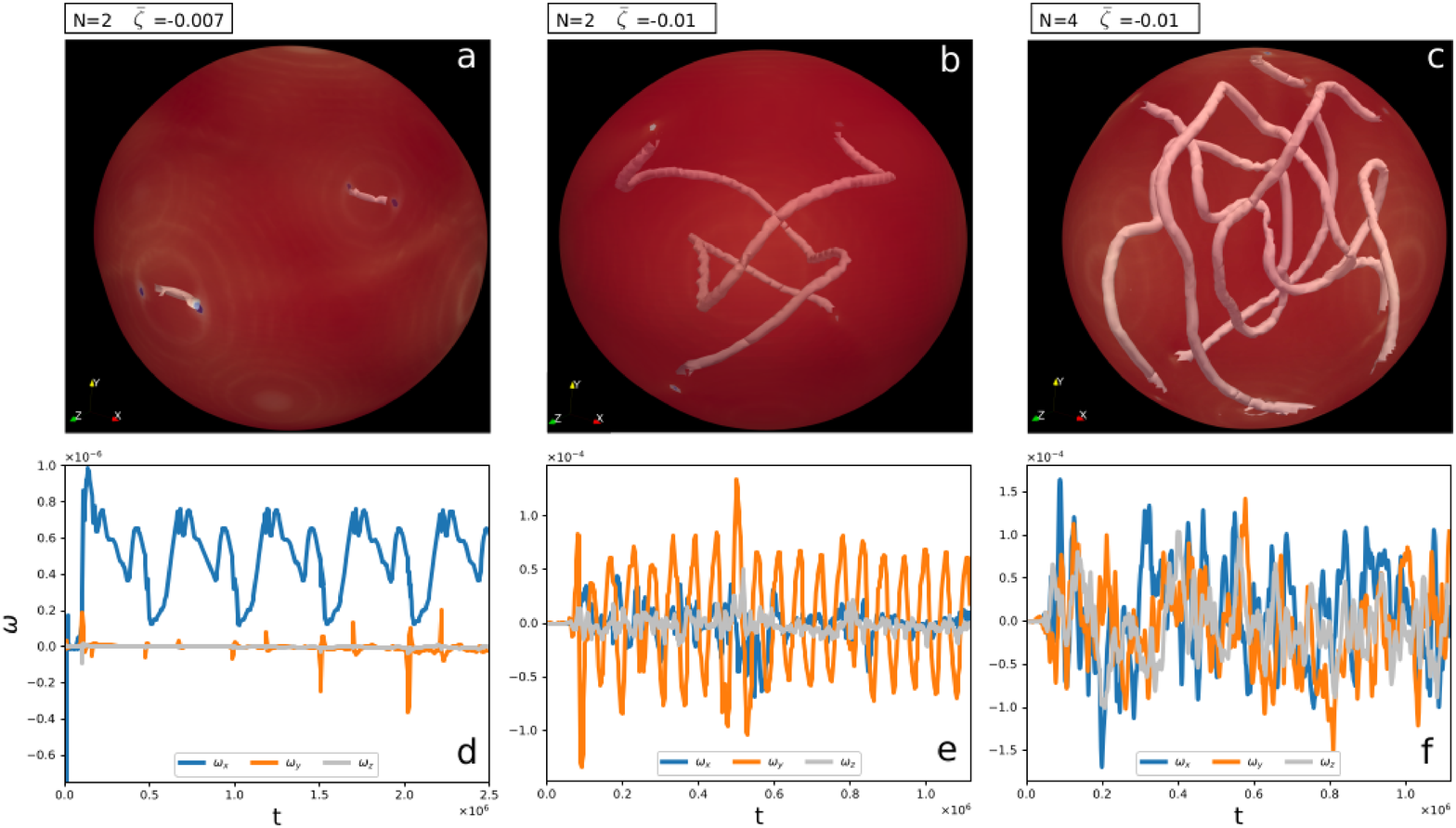}}
\caption{\textit{Active torque dipoles.} Panel a shows  a snapshot of the droplet and the disclination lines for the case at $N=2$ and $\bar{\zeta}=-7\times 10^{-3}$. In this case the droplet sets into rotational motion (notice the difference of the order of magnitude of the angular velocity in panel d with respect to the analogue cases presented in the main text for a droplet fueled by force dipoles only). Panel b and e show the case at $N=2$ and $\bar{\zeta}=- 10^{-2}$, characterized by the dancing of the disclination lines. Panel c shows  a snapshot of the droplet and its disclination lines for the case at $N=4$ and $\bar{\zeta}=- 10^{-2}$, in the chaotic regime -- see panel f -- characterized by nucleation of surface defects (panel c).}
\label{fig_suppl_1}
\end{figure*}

As discussed in the main text, the scenario in this case  is highly sensitive both at varying the intensity of active doping and the cholesteric power of the LC.  
The disclinations dance, described in Fig.~3 of the main text for  $N=1$, survives at higher $N$ only for sufficiently high values of  $|\bar\zeta|$. Our simulations confirm, that the range of stability of stationary rotation widens as $N$ is increased, while the set up of the disclination dance regime moves towards more intense $|\bar{\zeta}|$ (Fig.~\ref{fig_suppl_1}a,d,b,e). Moreover, if active doping exceeds a critical threshold, $|\bar{\zeta}| > 12 \times 10^{-3}$, nucleation of further defects (see Fig.~\ref{fig_suppl_1}c) on the droplet surface leads to droplet deformation with consequent chaotic dynamics~(Fig.~\ref{fig_suppl_1} f). Moreover, if $N \geqslant 4 $, the droplet directly moves from the rotational to the chaotic regime, analogously to what happens in a cholesteric droplet fueled by force dipoles only (see phase diagram in Fig.~2i of the main text).


\section*{Methods}

The dynamical equations have been integrated by means of a  hybrid lattice Boltzmann (LB) method~\cite{denniston2001,cates2009}, where the hydrodynamics is solved through a \emph{predictor-corrector} LB algorithm, while the dynamics of the order parameter has been treated with a finite-difference approach implementing a first-order upwind scheme and fourth-order accurate stencils for the computation of spacial derivatives.
The numerical code has been parallelized by means of Message Passage Interface (MPI), by dividing the computational domain in slices and
by implementing the ghost-cell method to compute derivatives on the boundary of the computational subdomains.
Runs have been performed using $24$, $32$ or $64$ CPUs in three-dimensional geometries, on a computational box of size $128^3$ or $100^3$, for at least $2 \times 10^6$ lattice Boltzmann iterations (corresponding to $\sim 12d$ of CPU-time). Periodic boundary conditions have been imposed. The director field has been randomly initialized inside the droplet and posed to $0$ outside.

The strain-rotational derivative appearing in the Beris-Edwards equation (Eq.~[3] of the main text)   accounts for the advective and orientational response of the LC to the flow and its explicit expression is given by:
\begin{multline}
\mathbf{S}(\mathbf{W},\mathbf{Q}) = (\xi \mathbf{D} + \mathbf{\Omega})(\mathbf{Q}+\mathbf{I}/3) + (\mathbf{Q}+\mathbf{I}/3)(\xi \mathbf{D}  - \mathbf{\Omega})  \\
- 2 \xi (\mathbf{Q}+\mathbf{I}/3) Tr (\mathbf{Q}\mathbf{W}),
\label{eqn:material_derivative}
\end{multline}
with $\mathbf{D}$ and $\mathbf{\Omega}$ respectively denoting the symmetric and asymmetric part of the velocity gradient tensor $\mathbf{W}= \mathbf{\nabla v}$. The parameter $\xi$ controls the aspect-ratio of the liquid crystal molecules and aligning properties to the flow (we chose $\xi=0.7$ to consider flow-aligning rod-like molecules).

The passive stress tensor, accounting for the dissipative/reactive contributions to the Navier-Stokes equation (Eq.~[4] of the main text), can be expressed as the sum of the isotropic pressure $\sigma_{\alpha \beta}^{hydro}= -p \delta_{\alpha \beta}$, with $p$ the hydrodynamic pressure, and the viscous stress $\sigma^{visc}_{\alpha \beta} = 2 \eta D_{\alpha \beta}$, with $\eta$ the shear viscosity.
The dynamics of the two order parameters $\phi$ and $\mathbf{Q}$ affect the hydrodynamics through the following passive terms:
\begin{equation}
\sigma^{bm} = \left(f-\dfrac{\delta\mathcal{F}}{\delta \phi} \right)\delta_{\alpha \beta} - \dfrac{\delta \mathcal{F}}{\delta (\partial_\beta \phi)} \partial_\alpha \phi,
\end{equation}
where $f$ is the free energy density, 
\begin{multline}
\sigma_{\alpha \beta}^{el} = -\xi H_{\alpha \gamma} \left(Q_{\gamma \beta} + \dfrac{1}{3} \delta_{\gamma \beta} \right) -\xi  \left(Q_{\alpha \gamma} + \dfrac{1}{3} \delta_{\alpha \gamma} \right) H_{\gamma \beta} \\	 + 2\xi \left(Q_{\alpha \beta} - \dfrac{1}{3} \delta_{\alpha \beta} \right) Q_{\gamma \mu} H_{\gamma \mu} + Q_{\alpha \gamma} H_{\gamma \beta}  - H_{ \alpha \gamma} Q_{\gamma \beta} .
\end{multline}

The values of free energy parameters are $a=0.07$, $k_{\phi}=0.14$, $A_0=1$, $K_Q=0.01$, and $W=0.02$. The rotational diffusion constant $\Gamma$ is set to  
$2.5$, while the diffusion constant to $M=0.1$.
All physical observables have been reported in lattice units, as usual in computational works on active matter.

The angular velocity of the droplet has been computed as:
$ \omega = \int \text{d}\mathbf{r} \phi \dfrac{\Delta \mathbf{r} \times \Delta \mathbf{v}}{|\Delta \mathbf{r}|^2},$ where $\Delta \mathbf{r}= \mathbf{r} - \mathbf{R}$ and $\Delta \mathbf{v}= \mathbf{v} - \mathbf{V}$, being $\mathbf{R}$ and $\mathbf{V}$ respectively the position and the velocity of the center of mass of the droplet. 
The degree of biaxiality of the LC has been computed by following the approach of Ref.~\cite{callan2006} as the second parameter of the Westin metrics $c_p = 2(\tilde\lambda_2-\tilde\lambda_3)$, 
where $\tilde\lambda_1$, $\tilde\lambda_2$ and $\tilde\lambda_3$ (with $\tilde\lambda_1\ge\tilde\lambda_2\ge\tilde\lambda_3$) are three eigenvalues of the positive definite diagonalised matrix $G_{\alpha\beta}=Q_{\alpha\beta}+\delta_{\alpha\beta}/3$.

\begin{table}[b]
\centering
{\begin{tabular}{@{}|l|l|l|@{}}
\hline
\multicolumn{1}{|c|}{\cellcolor{white}{\color{black}Model parameters }}   &
\multicolumn{1}{|c|}{\cellcolor{white}{\color{black}Simulation units}}   &
\multicolumn{1}{|c|}{\cellcolor{white}{\color{black}Physical units}}   \\\hline
Shear viscosity, $\eta$           & $5/3$  & $1.67 \ \text{KPas}$       \\
Elastic constant, $K_Q$           & $0.01$  & $50 \ \text{nN}$      \\
Shape factor, $\xi$               & $0.7$  & dimensionless       \\
Diffusion constant, $D=Ma$        & $0.007$  & $0.06\ \text{$\mu$} \text{m}^2 \text{s}^{-1}$     \\

Activity, $\zeta$    & $0-0.002$  & $(0-20) \ \text{KPa}$       \\ 
\hline
\end{tabular}\caption{Mapping of some relevant quantities between simulations units and physical units.}\label{tabel_units}}
\end{table}

\section*{Mapping with physical units}
By following previous studies~\cite{bonelli2019,tjhung2012}, an approximate relation between simulation and physical units (for an active gel of cytokeletal extracts) can be obtained using as length-scale, time-scale and force-scale respectively $L=1\mu m$, $\tau=10ms$ and $F=1000nN$. A mapping of some relevant quantities is reported in Table~\ref{tabel_units}.

\end{document}